\documentclass{pasj00}
\draft

\begin{document}
\SetRunningHead{Author(s) in page-head}{Running Head}

\title{Location of gamma-ray flaring region in quasar 4C +21.35}


\author{Maichang Lei}
\affil{Yunnan Observatory, Chinese Academy of Sciences,  Kunming 650011, China}
\email{maichanglei83@ynao.ac.cn}

\author{Jiancheng Wang}
\affil{Yunnan Observatory, Chinese Academy of Sciences,  Kunming 650011, China}
\email{jcwang@ynao.ac.cn}

%

\KeyWords{galaxies: active - gamma rays: general - quasars: individual (4C +21.35) - radiation mechanism: non-thermal}

\maketitle

\begin{abstract}
4C +21.35 is a flat-spectrum-radio-quasar-type blazar, in which the rapid variability of very high energy (VHE, $E_{\gamma}\gtrsim 100$\,GeV) emission as short as $\sim$ 10 minutes was observed by MAGIC Cherenkov telescopes, and the VHE spectrum extends up to at least 400\,GeV.
In this paper, by using a flat broad-line region (BLR) structure, we study the location and  properties of $\gamma$-ray emitting region
of 4C +21.35 under the constraints of multiwavelength data.
We fit three quasi-simultaneous spectral energy distributions (SEDs) using homogeneous one-zone leptonic model, in which the flat BLR with the aperture angle of $\alpha=25^{\circ}$ and a spherically symmetric hot dusty torus with the temperature of $T_{\rm sub}=1200$\,K, are assumed.
The results show that the jet structure of 4C +21.35 is nearly conical with a half-opening angle of $\theta_{\rm j}\simeq 0.29^{\circ}-0.6^{\circ}$. Furthermore, the emitting region is located within the BLR clouds and approaches to outer radius of the BLR during the flaring states, while it is well beyond the dusty torus in quiescent state. The quiescent high-energy emission is dominated by synchrotron self-Compton (SSC) process, the high-energy emission during the flaring periods is dominated by Compton scattering of BLR and dusty torus photons with the value of Compton-dominance parameter is about 30. Moreover, the fit to optical/ultraviolet data provides a further support that the central black hole (BH) mass of 4C +21.35 is $6\times 10^{8}$\,$M_{\odot}$.
\end{abstract}

\section{Introduction} \label{1. Introduction}
Only $\sim 10-20\%$ radio-loud active galactic nuclei (AGNs) have jet close to our line of sight, i.e. blazars \citep{2011AIPC.1381..180G}.
Their emissions are dominated by non-thermal emission throughout entire electromagnetic spectrum.
Blazars are an outstanding subclass of AGNs, which consists of BL Lacertae objects (BL Lacs) and flat spectrum radio quasars (FSRQs) \citep{1995PASP..107..803U}.

BL Lacs and FSRQs have significant differences in their structures. BL Lacs could have a radiatively inefficient accretion disk and BLR \citep{2011ApJ...732..113S}, but lack of dusty torus. Instead, FSRQs have a radiatively efficient accretion disk, together with BLR and dusty torus. The emission from BL Lacs is generally dominated by synchrotron self-Compton (SSC) process \citep{2008ApJ...686..181F,2009ApJ...692...32D}, while the emission from FSRQs is dominated by the external radiation Compton (ERC) mechanism, in which the seed photons could originate from the accretion disk \citep{1992A&A...256L..27D,1993ApJ...416..458D}, BLR \citep{1994ApJ...421..153S,2003APh....18..377D,2006ApJ...653.1089L,2007MNRAS.375..417C,2008MNRAS.387.1669G,2009MNRAS.397..985G,2012arXiv1209.2291T,2014PASJ...66....7L}
 and dusty torus \citep{2003APh....18..377D,2000ApJ...545..107B} radiation fields.

4C +21.35 (PKS 1222+216, z=0.432; \cite{1987ApJ...323..108O}) is a $\gamma$-ray-emitting FSRQ \citep{1999ApJS..123...79H}, confirmed by the Energetic Gamma Ray Experiment Telescopes (EGRET). During the first three months of scientific operation of the Fermi Large Area Telescope (Fermi-LAT), the flux of this source was too low to be included in the Fermi-LAT Bright Source List \citep{2009ApJS..183...46A}. Fortunately, it was included in the First LAT Catalog \citep{2010ApJS..188..405A}, which was associated with 1FGL J1224.7+2121. Fermi-LAT observations show that 4C +21.35 was in a quiescent state during 2008 August to 2009 September and subsequently the source became active. On 2010 April 24, the Fermi-LAT detected a strong outburst from this source \citep{2010ATel.2584....1D}. During this outburst, the observed $\gamma$-ray emission extends up to the photon energies larger than 100\,GeV
\citep{2010ATel.2617....1N,2011A&A...529A..59N}. In 2010 April and June, the source arose a strong GeV outburst, several major flares are described by rise and decay timescales of order of day \citep{2011ApJ...733...19T}. On 2010 June 17, the Major Atmospheric Gamma Imaging Cherenkov Telescope (MAGIC) detected $\gamma$-ray emission from 4C +21.35, where the flux-doubling timescale of the $\gamma$-ray was about 10 minutes, and the spectrum extends from about 70\,GeV up to at least 400\,GeV \citep{2011ApJ...730L...8A}.

Currently, several explanations on the production of the short $\gamma$-ray variability of 4C +21.35 have been put forward. \citet{2014MNRAS.442..131K} considered the emitting region to be at the recollimation zone of the jet. \citet{2011A&A...534A..86T} proposed three scenarios to explain the SEDs and favored the emitting region to be re-collimated and focussed zone of the jet at larger distance from the black hole (BH).
In addition, photons/axion transition was proposed to explain fast $\gamma$-ray variability \citep{2012PhRvD..86h5036T, 2014arXiv1409.4401G}. The comprehensive SED investigations to 4C +21.35 have been performed by \citet{2014ApJ...786..157A} (hereafter, Paper-I), in which three quasi-simultaneous SEDs from 2009 to 2011 were modeled with a combination of SSC and Compton-scattered soft photons from dusty torus. In addition, \citet{2012ApJ...755..147D} argued that the rapid VHE radiation originates from a beam of leptonic secondaries initiated by the ultra-high energy cosmic rays (UHECRs) at the pc-scale region.

For 4C +21.35, the very rapid $\gamma$-ray variability indicates an extremely compact emitting region, with size $R_{\rm b}^{\prime}\leq ct_{\rm var}\delta_{\rm D}/(1+z)\simeq 1.26\times 10^{15}(\delta_{\rm D}/100)(t_{\rm var}/10\,\rm minutes)$\,cm, where $t_{\rm var}$ is the observed flux-doubling timescale and $\delta_{\rm D}$ is the Doppler factor. Here, we have assumed that the emitting region has a higher Doppler factor of 100. Under conical configuration, the half-opening angle of relativistic jet is generally assumed to be $\sim$0.1, this corresponds to the emitting region at radial distance $r\sim 10^{16}$\,cm. Generally, the BLR has the scale of $\sim 0.1-1$\,pc, this implies that the emitting region is well within the BLR cavity. Under this situation, the low-energy photons originating from BLR are the target source for ERC process. Meanwhile, they will also severely absorb the $\gamma$-rays above a few tens of GeV via $e^{\pm}$ pair production due to $\gamma-\gamma$ interactions \citep{2006ApJ...653.1089L}. The use of a flat BLR will effectively alleviate this absorption effect, as presented by \citet{2014PASJ...66....7L}.

In Section 2, we give a brief description of the model. In Section 3, we perform spectral modeling to three SED data points. Our discussion is given in Section 4. The conclusion is given in Section 5. Symbols with a numerical subscript should be read as a dimensionless number $X_{\rm n}=X/(10^{\rm n}\,\rm cgs~units)$.
Throughout the paper, unprimed quantities are measured in the observer's frame, unless stated otherwise, and primed quantities refer to the blob's frame, while the starred quantities refer to the stationary frame with respect to the BH.
We adopt the cosmological parameters: $H_{0}=71$\,km~s$^{-1}$~Mpc$^{-1}$, $\Omega_{\rm m}=0.3$, $\Omega_{\Lambda}=0.7$.

\section{Model description}
The model is set up within the framework of homogeneous one-zone leptonic model, in which for simplicity the emitting region is assumed to be a spherical blob with the
bulk Lorentz factor $\Gamma_{\rm j}$. The jet makes an angle $\theta_{\rm v}$ with respect to the line of sight. The blob is filled by tangled, homogeneous magnetic field of strength $B$, together with relativistic electrons (including electrons and positrons).
The relativistic effects are characterized by Doppler factor $\delta_{\rm D}=[\Gamma_{\rm j}(1-\beta_{\Gamma}\cos\theta_{\rm v})]^{-1}$, where $\beta_{\Gamma}$ is the bulk velocity in units of the speed of light c. The comoving size of the blob $R_{\rm b}^{\prime}$ is related to the observed variability timescale $t_{\rm var}$ through $R_{\rm b}^{\prime}\simeq t_{\rm var} {c\delta_{\rm D}}/{(1+z)}$.

A broken power-law function with a cutoff is adopted to describe the relativistic electron distribution within the emitting region:
\begin{eqnarray}
n_{\rm e}^{\prime}(\gamma^{\prime}) &=& \frac{n_{0}{\gamma_{\rm b}^{\prime}}^{-1}}{(\gamma^{\prime}/\gamma_{\rm b}^{\prime})^{s_{1}}+(\gamma^{\prime}/\gamma_{\rm b}^{\prime})^{s_{2}}},
\end{eqnarray}
where $n_{0}$ is the normalization factor; $\gamma^{\prime}=E_{\rm e}^{\prime}/m_{\rm e}c^{2}$ is the electron Lorentz factor assumed to vary within the range $\gamma_{\rm min}^{\prime}\leq \gamma^{\prime} \leq \gamma_{\rm max}^{\prime}$. $s_{1}$ and $s_{2}$ are the spectral indices below and above the break energy $\gamma_{\rm b}^{\prime}$, respectively.

An important parameter to characterize 4C +21.35 is the central BH mass, which is still a matter of debate. \citet{2004ApJ...615L...9W} suggested a value of $1.48\times10^{8}$\,$M_{\odot}$ obtained by the relation between $\rm H\beta$ broad line and the continuum luminosity-BLR radius. Recently,
based on the virial method from the full width at half-maximum of $\rm M_{\rm gII}$, $\rm H\beta$ and $\rm H\alpha$ broad lines and thermal continuum luminosity,
\citet{2012MNRAS.424..393F} presented a larger value of $\sim 6\times 10^{8}$\,$M_{\odot}$, which agrees well with the values found by \citet{2011ApJS..194...45S} and \citet{2012ApJ...748...49S}. Their results originate from Sloan Digital Sky Survey (SDSS) spectra. Given these results, the BH mass of 4C +21.35 $M_{\rm BH}=6\times 10^{8}$\,$M_{\odot}$ is adopted in our model. Beyond the BH, there is an optically thick accretion disk; its spectral luminosity is assumed to be characterized by a Shakura-Sunyaev disk \citep{1973A&A....24..337S} spectrum in the form \citep{2014ApJ...782...82D}
\begin{eqnarray}
\epsilon_{\star} L_{\rm d}(\epsilon_{\star})&=& 1.12L_{\rm d}~\biggr(\frac{\epsilon_{\star}}{\epsilon_{\rm max}}\biggr)^{4/3}\exp(-\epsilon_{\star}/\epsilon_{\rm max}),
\end{eqnarray}
where $\epsilon_{\star}=E_{\star}/m_{\rm e}c^{2}$ is the dimensionless photon energy in stationary frame, which is related to observed dimensionless photon energy by $\epsilon_{\star}=(1+z)\epsilon$. The value of $\epsilon_{\rm max}$ is the function of the spin of the BH and relative Eddington luminosity.

Generally, the BLR is assumed to be spherical, and with a radius that can be written in terms of the accretion disk luminosity $L_{\rm d}$ as \citep{2008MNRAS.387.1669G}
\begin{eqnarray}
R_{\rm BLR} &\simeq& 0.1\sqrt{\frac{L_{\rm d}}{10^{46}\,\rm erg~s^{-1}}}\,\rm pc.
\end{eqnarray}
In this paper, we consider a flat BLR, in which the cloud matter is filled within a region with width extending from $r_{\rm BLR,in}=(R_{\rm BLR}-0.05)$\,pc to $r_{\rm BLR,out}=(R_{\rm BLR}+0.05)$\,pc. A detailed description of the flat BLR structure is given by \citet{2014PASJ...66....7L}.

Moreover, the spectral luminosity of the infrared (IR) component can be approximated by a graybody radiation field, namely,
\begin{eqnarray}
\epsilon_{\star} L_{\rm IR}(\epsilon_{\star})&=& \frac{15L_{\rm IR}}{\pi^{4}}\frac{(\epsilon_{\star}/\Theta)^{4}}{\exp(\epsilon_{\star}/\Theta)-1},
\end{eqnarray}
where $\Theta=k_{\rm B}T_{\rm sub}/m_{\rm e}c^{2}$, the sublimation temperature of the dusty torus of 4C +21.35 is $T_{\rm sub}=1200$\,K \citep{2011ApJ...732..116M}. For simplicity, the dusty torus is assumed to have a geometrically thin and spherically symmetric structure with radius that corresponds to dust sublimation \citep{2011ApJ...732..116M}, i.e.,
\begin{eqnarray}
R_{\rm torus}&\simeq& 0.4\sqrt{\frac{L_{\rm d}}{10^{45}\,\rm erg~s^{-1}}}\biggr(\frac{T_{\rm sub}}{1500\,\rm K}\biggr)^{-2.6}~\rm pc.
\end{eqnarray}

Combined the $\rm H\beta$ luminosity, $L_{\rm H\beta}\simeq 2\times 10^{43}$\,erg~s$^{-1}$, and the scaling relation $L_{\rm BLR}\simeq 25.3~L_{\rm H\beta}$ \citep{2004ApJ...615L...9W,2006ApJ...646....8F}, \citet{2011ApJ...733...19T} derived the accretion-disk luminosity as $L_{\rm d}\simeq 10~L_{\rm BLR}\simeq 5\times 10^{45}$ \,erg~s$^{-1}$, where the fraction 0.1 of the disk radiation reprocessed by BLR clouds has been assumed. \citet{2011A&A...534A..86T} assumed that the Swift ultraviolet spectrum of 4C +21.35 is contributed by the accretion-disk radiation, and estimated that $L_{\rm d}\simeq 5\times 10^{46}$\,erg~s$^{-1}$. Moreover, \citet{2014ApJ...786..157A} performed the SED fitting to 4C +21.35, and restricted the accretion-disk luminosity to $L_{\rm d}=0.2~L_{\rm Edd}=1.56\times 10^{46}$\,erg~s$^{-1}$. We use the larger disk luminosity here, noting that a larger $L_{\rm d}$ corresponds to a larger BLR radius, subsequently reducing the absorption efficiency of the $\gamma$-rays.

\section{VHE emission production and the SED modeling }
\subsection{Constraint on the site of the VHE $\gamma$-ray production}
According to the aforementioned discussions, the $\gamma$-ray emitting region associated with $\sim10$\,minutes VHE variability is roughly located at $\simeq 1.3\times10^{16}$\,cm $\simeq$ $5\times 10^{-3}$\,pc, assuming that this region takes up entire cross-section of the jet. In a spherical BLR scenario, the $\gamma$-rays with energies above tens of GeV are expected to be strongly attenuated by the interactions with low-energy photons from the BLR.
Therefore, it is useful to probe the location of the $\gamma$-ray emitting region based on a flat BLR structure.

\begin{figure}
  \centerline{
  \FigureFile(120mm,80mm){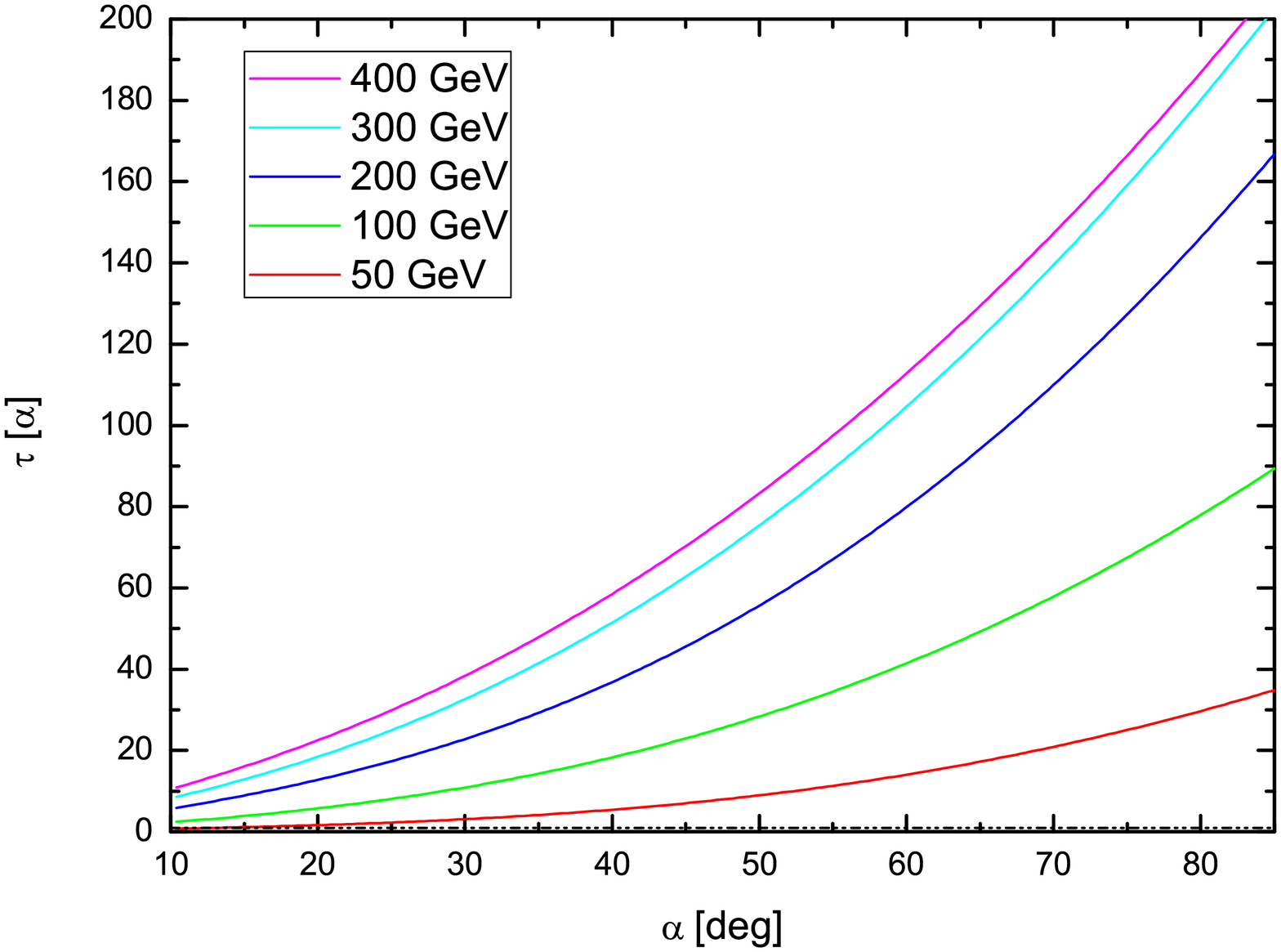}}
  \caption{Pair-production opacity $\tau$ versus the aperture angle $\alpha$ of the BLR for different $\gamma$-ray energies from $E_{\gamma}=50$\,GeV upward to $E_{\gamma}=400$\,GeV. The location of the $\gamma$-ray emitting region is fixed at $r_{\rm b}=5\times 10^{-3}$\,pc from BH, which corresponds roughly to 10\,minutes VHE $\gamma$-ray variability. The accretion disk luminosity is $L_{\rm d}=5\times 10^{46}$\,erg~s$^{-1}$. The horizonal dashed line corresponds to $\tau=1$.}
\label{fig:Colored-tau}
\end{figure}

Figure \ref{fig:Colored-tau} illustrates a calculation of the pair-production opacity for different $\gamma$-ray energies from $E_{\gamma}=50$\,GeV up to $E_{\gamma}=400$\,GeV, i.e., throughout the whole electromagnetic spectrum available by MAGIC from 4C +21.35. \citet{2011ApJ...730L...8A} have shown that the de-absorbed VHE spectrum smoothly connects to the Fermi-LAT spectrum, suggesting that the same population is powering both emissions. Generally, the short variability is thought to be related to the compact emission region, and correspondingly with shorter distance from the BH. In order
to explore the location of fast VHE variability observed on 2010 June 17,
the location of the emitting region is fixed at $r_{\rm b}=5\times 10^{-3}$\,pc, corresponding roughly to the short timescale variability of 10\,minutes. However, it should be emphasized here that in comparison to the
observation on 2010 June 17, Fermi-LAT observations during its active period showed that the source generated a few $\gamma$-rays flares in the MeV-GeV band with longer
variability timescales \citep{2011arXiv1110.4471F,2014ApJ...796...61K}.

We can see from Figure \ref{fig:Colored-tau} that the observed VHE photons can not be produced within the BLR cavity. This constraint is also effective for lower $\gamma$-rays with energy $E_{\gamma}=50$\,GeV, where the opacity $\tau$ remains larger than unity, even for $\alpha\simeq 10^{\circ}$. Therefore, if the jet of 4C +21.35 has a half-opening angle $\sim 0.1$, it is not appropriate to make the assumption that the location of emitting region can be determined by combining the measured variability timescale, $R_{\rm b}^{\prime}\simeq c\delta_{\rm D}t_{\rm var}$, and the assumption that the emitting region covers the entire cross-section of the jet. Under this case, the location of the emitting region related to $\sim10$\,minutes variability must be located within the BLR cavity, in direct conflict with the theoretical expectation. This seemly indicates that the jet structure of 4C +21.35 is exceptionally complicated, the ultra-short variability could originate from the spine of the structured jet, or from the re-collimated and focussed zone of the jet.

\begin{figure}
  \centerline{
  \FigureFile(120mm,80mm){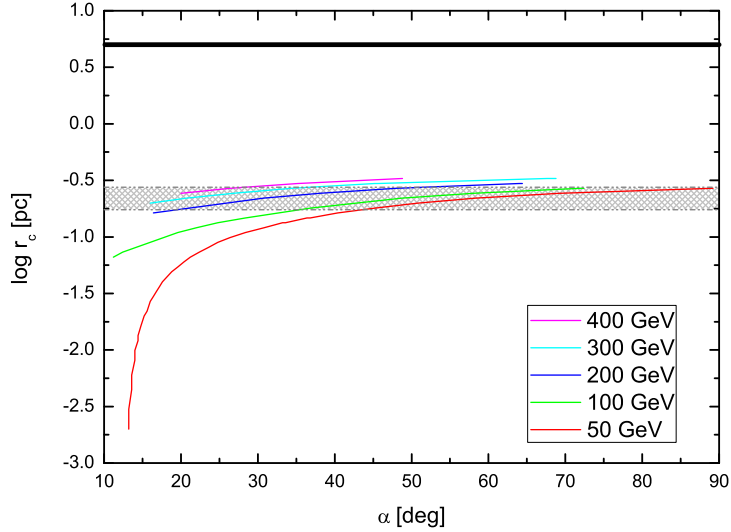}}
  \caption{The critical position $R_{\rm c}$ of the $\gamma$-ray emitting region versus the aperture angle $\alpha$ of the BLR for different $\gamma$-ray energies from $E_{\gamma}=50$\,GeV upward to $E_{\gamma}=400$\,GeV. The shaded region represents the one surrounded by inner radius $r_{\rm BLR,in}$ and outer radius $r_{\rm BLR,out}$ of the BLR. The accretion disk luminosity is $L_{\rm d}=5\times 10^{46}$\,erg~s$^{-1}$. The horizonal thick line represents the scale of the dusty torus.}
\label{fig:Colored-Rc}
\end{figure}

In our scheme, the pair-production opacity $\tau$ depends on the $\gamma$-ray energy $E_{\gamma}$ and on two geometrical parameters: the BLR aperture angle $\alpha$ and
the location of the emitting region $r$. Letting $\tau=1$, we can find out the critical position $r_{\rm c}$ of the emitting region, as shown in Figure \ref{fig:Colored-Rc}. The shaded region represents the part surrounded by inner radius $r_{\rm BLR,in}$ and outer radius $r_{\rm BLR,out}$ of the BLR, whereas the horizonal thick line represents the scale of the dusty torus. In calculation, the opacity $\tau$ is a path integration along the line of sight from $r_{\rm c}$ to $r_{\rm max}$, where $r_{\rm max}=5$\,pc is fixed. It is clear that when the emitting region is within the BLR cavity, just for $\alpha$ less than $35^{\circ}$, the $\gamma$-rays with energy $E_{\gamma}=100$\,GeV can escape the absorption of the BLR photons. For the observed higher-energy $\gamma$-ray photons, the location of the emitting region must be within or beyond the BLR clouds, this provides strong constraint on the location of $\gamma$-ray-emitting zones for FSRQs.

\subsection{Modeling the SEDs of 4C +21.35}
\begin{figure*}
\includegraphics[scale=0.3,angle=0]{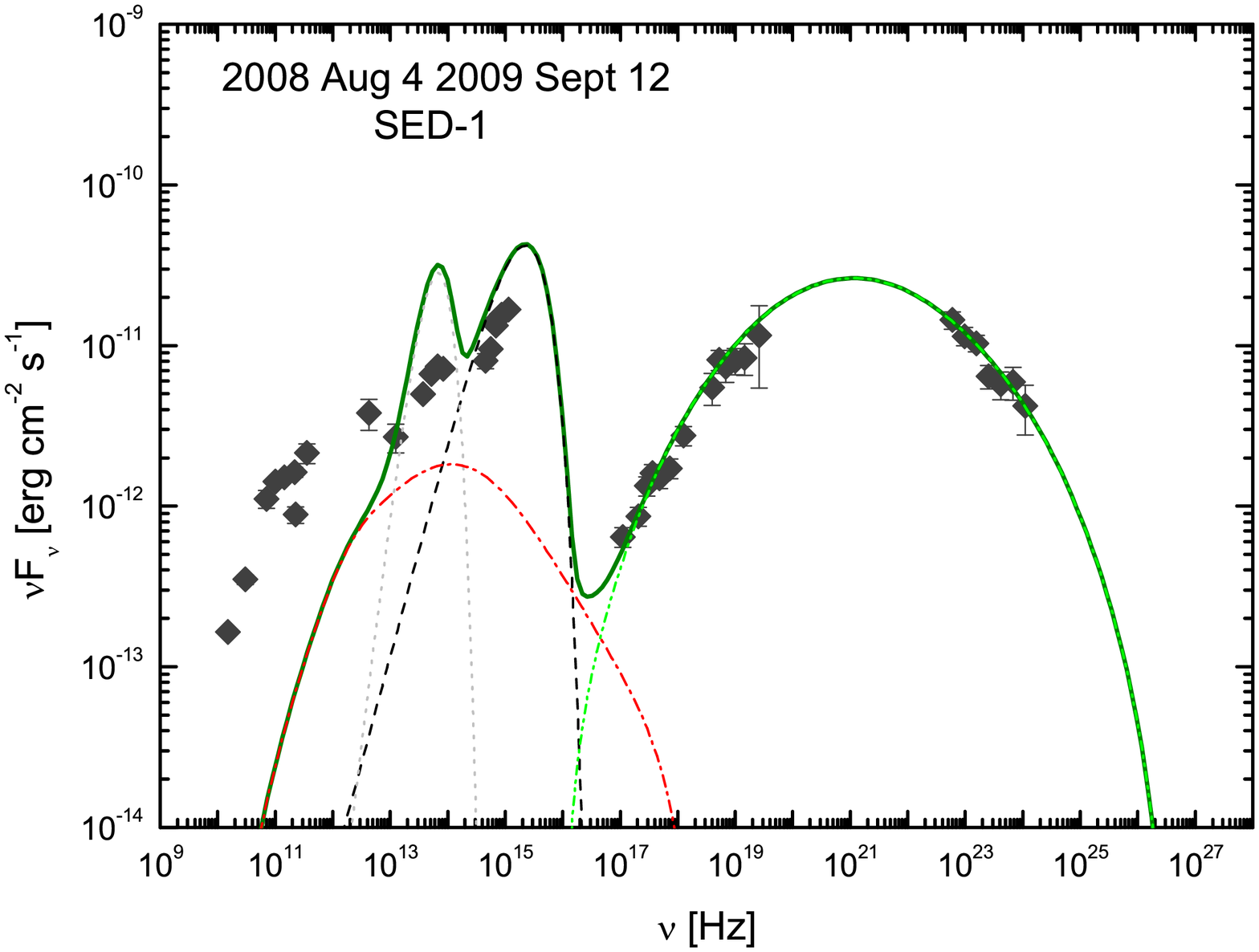}
\includegraphics[scale=0.3,angle=0]{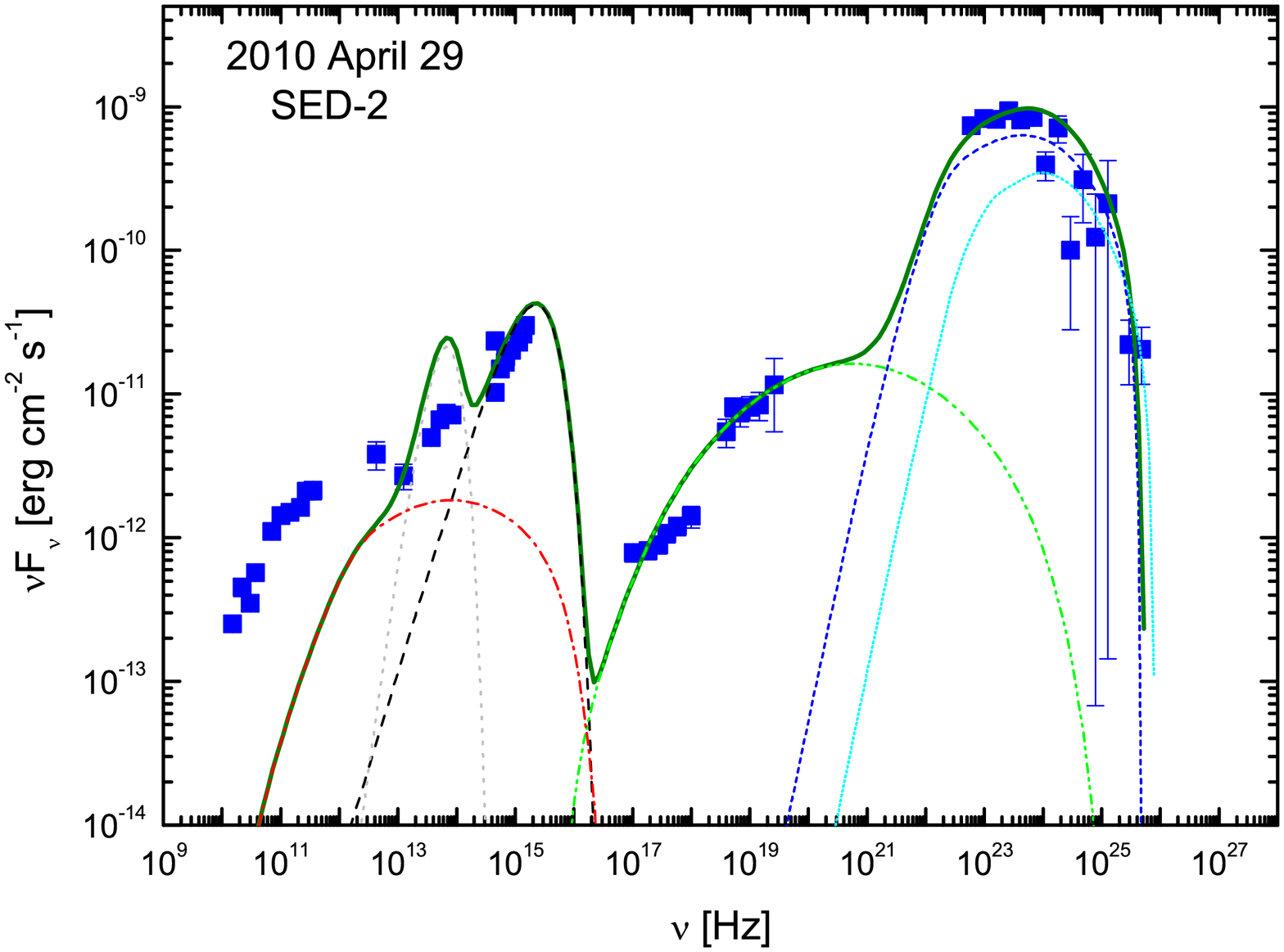}\\
\includegraphics[scale=0.3,angle=0]{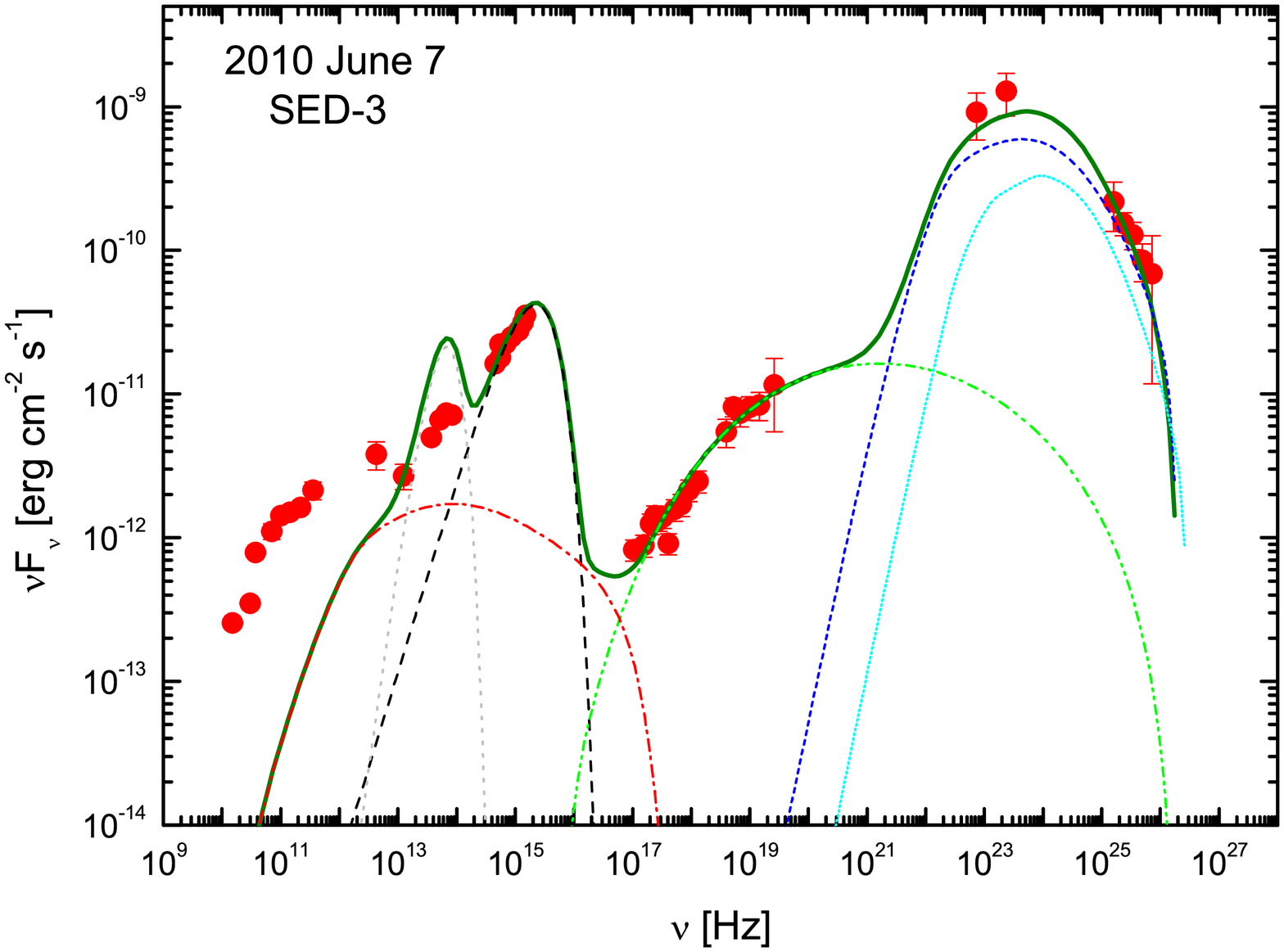}
\caption{Spectral energy distributions of 4C +21.35 in three epochs: 2008 August 4 to 2009 September 12 (dark-gray diamonds), 2010 April 29 (blue squares) and 2010 June 17 (red circles).
Dashed lines indicate synchrotron, the dusty torus, the accretion disk and SSC emission components, as well as the Compton-scattered the dusty torus and the BLR emissions, respectively. The thicker solid line is the sum of all the components. The MAGIC data have been corrected for the EBL absorption using the model of \citet{2010ApJ...712..238F}.}
\label{fig:SED-111}
\end{figure*}

Three quasi-simultaneous SEDs shown in Figure \ref{fig:SED-111}, were presented originally by Paper-I. The MAGIC data have been corrected for EBL absorption using the model of \citet{2010ApJ...712..238F}.
These SEDs include a quiescent state, in which the temporal flux is integrated over time spanning from 2008 August 4 to September 12 (dark-gray diamonds, SED-1), together with the flaring states of 2010 April 29 (blue squares: SED-2) and 2010 June 17 (red circles, SED-3).

We model the SEDs of three epochs within the framework of one-zone leptonic scenario. At first, we obtain information on the peaks of the spectrum and locations at large. In synchrotron hump, three peaks could occur, including the first peak at radio band, the second one at infrared band from the reprocessed emission of the dusty torus, and the third one at ultraviolet band from the emission of the accretion disk. In the second hump, we can not identify the number of peak and the corresponding location explicitly. However, two peaks could appear at MeV and GeV bands, in which the first peak could from the SSC emission and/or the Compton scattering of low-energy photons from accretion disk depending on the distance from the BH, and the second peak could originate from the Compton scattering of the diffuse photons from the BLR and/or the dusty torus. In the SED modeling, we due with the radio data as an upper limits, because they could be the superposition of multiple self-absorbed components or originate from the extended regions of relativistic jet.

In order to reduce the number of free parameters used in the model, several parameters have been fixed based on the following arguments:

(1) The sizes of the emitting regions for three multiwavelength campaigns are equally fixed at $R_{\rm b}^{\prime}=1\times 10^{15}$\,cm, which corresponds to short variability with timescale $\sim 10$\, minutes.

(2) The measurement of the apparent superluminal motion to relativistic jet of 4C +21.35 gives a bulk Lorentz factor $\Gamma_{\rm j}\approx 20$ \citep{2001ApJ...556..738J} on parsec scales, and constrains the opening angle within $\sim 3^{\circ}$. Moreover, blazar jets have the opening angles that satisfy the relation $\theta_{\rm j}\sim (0.1-0.2)/\Gamma_{\rm j}$ \citep{2009A&A...507L..33P,2013A&A...558A.144C} on average. Here we use a larger jet half-opening angle of $\theta_{\rm j}\simeq 0.2/\Gamma_{\rm j}$, limiting the jet opening angle of 4C +21.35 to be $\theta_{\rm j}\sim 0.57^{\circ}$. In SED modeling, we take the value $\theta_{\rm j}=0.6^{\circ}$ approximately, and the viewing angle is equal to the jet half-opening angle, $\theta_{\rm v}=\theta_{\rm j}$.

(3) The typical BLR thickness ratio was found to be $\rm H/R \sim 0.4$ \citep{2008MNRAS.387.1237D,2011MNRAS.413...39D,2012ApJ...748...49S},  corresponding to $\alpha\simeq 25^{\circ}$ approximately. Thus we fix the aperture angle $\alpha$ of the BLR to be $25^{\circ}$. We also assume that the BLR clouds intercept a fraction $f_{\rm BLR}=0.1$ of the central continuum into the diffuse radiation and the dusty torus intercepts a fraction $f_{\rm IR}=0.3$ of the one into the infrared. Again, we further assume that the BLR reprocesses a fraction of $f_{\rm cont}=0.05$ and $f_{\rm line}=0.05$ into continuum and emission lines, respectively.

We fit the ultraviolet data with a blackbody spectrum from the standard thin disk, in which the accretion-disk luminosity is $L_{\rm d}=4.6\times 10^{46}$\,erg~s$^{-1}$. Indeed, the infrared radiation of dusty torus is determined by the relation $L_{\rm IR}=f_{\rm IR}L_{\rm d}$. The BLR radiation field relies on the radial distance of the emitting region from the BH, thus we can relate $R_{\rm b}^{\prime}$ to the location of the emitting region via $r_{\rm e}\simeq R_{\rm b}^{\prime}/\tan\theta_{\rm j}$. Since $r_{\rm e}\simeq 3\times 10^{-2}$\,pc which is less than the scale of the BLR, the VHE photons will be absorbed by BLR photons. Thus, we must adjust the radial distance to avoid the absorption of VHE photons, and find that the radial distance $r_{\rm b}=8r_{\rm e}$ satisfies the SED modeling well.

The modeling SEDs are shown in Figure \ref{fig:SED-111}, and the parameters are listed in Table \ref{tab:tab-1} and Table \ref{tab:tab-2}, in which the cold proton power is estimated by assuming one cold proton per relativistic electron. It should be emphasized that the radiative power given in Table \ref{tab:tab-2} is absolute, which is related to apparent one by (see Appendix in \cite{2012ApJ...755..147D})
\begin{eqnarray}
L_{\rm abs,syn/SSC}&=& 8\Gamma_{\rm j}^{2}L_{\rm syn/SSC}/3\delta_{\rm D}^{4},
\nonumber\\
L_{\rm abs,EC}&=& 32\Gamma_{\rm j}^{4}L_{\rm EC}/5\delta_{\rm D}^{6}.
\end{eqnarray}

It is noted from Figure \ref{fig:SED-111} that the low-state SED from 2008 August 4 to September 12 is well reproduced by the SSC model, showing no hint of Compton scattering of BLR and/or dusty torus photons. This indicates that the location of the emitting region in the quiescent state is well beyond the dusty torus, at least $r>R_{\rm torus}\simeq 4.8$\,pc. As for both flaring states, the SEDs show significant Compton dominance, in which the parameter $A_{\rm C}$ ($\simeq L_{\rm IC}^{\rm iso}/L_{\rm syn}^{\rm iso}$, where $L_{\rm IC}^{\rm iso}$ and $L_{\rm syn}^{\rm iso}$ are isotropic Compton and synchrotron luminosity) are 29.5 and 28.5 for SED-2 and SED-3, respectively. While the ratios of isotropic SSC and synchrotron radiative luminosity are 11.4 and 12.4 during the flaring periods. The SED is comprised by Compton scattering of BLR and dusty torus radiation fields, which have comparable flux levels.
Table \ref{tab:tab-1} and Table \ref{tab:tab-2} show that both flaring states have almost the same parameters, implying that they produce radiation under roughly similar circumstance.

\subsection{Comoving energy density of the radiation fields}
Because the locations of the emitting regions of 4C +21.35 are within the BLR clouds during the flaring states, the cooling of relativistic electrons becomes more complicated. Here, we explore which cooling process dominates particle cooling in the emitting region.

The energy-loss rate of electrons in Thomson limit depends on the energy density in the comoving frame:
\begin{eqnarray}
u_{\rm ext}^{\prime} &=& \oint d\Omega^{\prime} \int_{0}^{\infty} d\epsilon^{\prime} (\frac{\epsilon^{\prime}}{\epsilon_{\star}})^{3}u(\epsilon_{\star},\Omega_{\star}).
\end{eqnarray}

Here, the photon energy $\epsilon_{\star}$ and angle $\theta_{\star}=\cos^{-1}\mu_{\star}$ in the stationary frame of the accretion disk are related to the comoving photon energy $\epsilon^{\prime}$ and direction cosine $\mu^{\prime}$ through the relations $\epsilon_{\star}=\Gamma_{\rm j} \epsilon^{\prime}(1+\beta_{\Gamma}\mu^{\prime})$, $\mu_{\star}=(\mu^{\prime}+\beta_{\Gamma})/(1+\beta_{\Gamma}\mu^{\prime})$, and $\phi_{\star}=\phi^{\prime}$.
Therefore, the comoving energy density of accretion disk is
\begin{eqnarray}
u_{\rm disc}^{\prime}(r) &=& \frac{3}{16}\frac{R_{\rm s}L_{\rm d}}{\pi c \eta \Gamma_{\rm j}^{4}} \int_{\mu_{\rm min}^{\prime}}^{\mu_{\rm max}^{\prime}}  \frac{d\mu^{\prime}}{(1+\beta_{\Gamma}\mu^{\prime})^{4}}\frac{\varphi(r)}{r^{3}},
\end{eqnarray}
where $c\beta_{\Gamma}=c/\sqrt{\Gamma_{\rm j}^{-2}-1}$ is the velocity of the moving plasma.

In contrast to the emission from accretion disk seen in the comoving frame, the emission from BLR will be Doppler-boosted, the total comoving energy density for emission lines is given by
\begin{eqnarray}
u_{\rm BLR,line}^{\prime}(r) &=& \frac{2\pi \Gamma_{\rm j}^{2}}{c} \sum_{i=1}^{m}\frac{N_{\epsilon_{\star},\rm i}}{N_{\epsilon_{\star},\rm tot}}
\int_{\theta_{\star,\rm min}}^{\theta_{\star,\rm max}}d\theta_{\star} \sin\theta_{\star}
\nonumber\\
\nonumber\\
&\times& I_{\rm line}(r,\theta_{\star})(1-\beta_{\Gamma}\cos\theta_{\star})^{2},
\end{eqnarray}
where m is the number of the emission lines. The total comoving energy density of the continuum is
\begin{eqnarray}
u_{\rm BLR,cont}^{\prime}(r) &=& \frac{2\pi \Gamma_{\rm j}^{2}}{c}\int_{\theta_{\star,\rm min}}^{\theta_{\star,\rm max}} d\theta_{\star}\sin\theta_{\star}
\nonumber\\
&\times& I_{\rm cont}(r,\theta_{\star})(1-\beta_{\Gamma}\cos\theta_{\star})^{2}.
\end{eqnarray}

At last, the total comoving energy density of the IR dusty torus can be easily written as
\begin{eqnarray}
u_{\rm IR}^{\prime}(r) &=& \frac{f_{\rm IR}L_{\rm d}}{24\pi R_{\rm IR}^{2}c\Gamma_{\rm j}^{4}} \Big[\frac{1}{(1+\beta_{\Gamma}\mu_{\rm min}^{\prime})^{3}}
\nonumber\\
&-& \frac{1}{(1+\beta_{\Gamma}\mu_{\rm max}^{\prime})^{3}}\Big].
\end{eqnarray}

\begin{figure}
  \centerline{
  \FigureFile(120mm,80mm){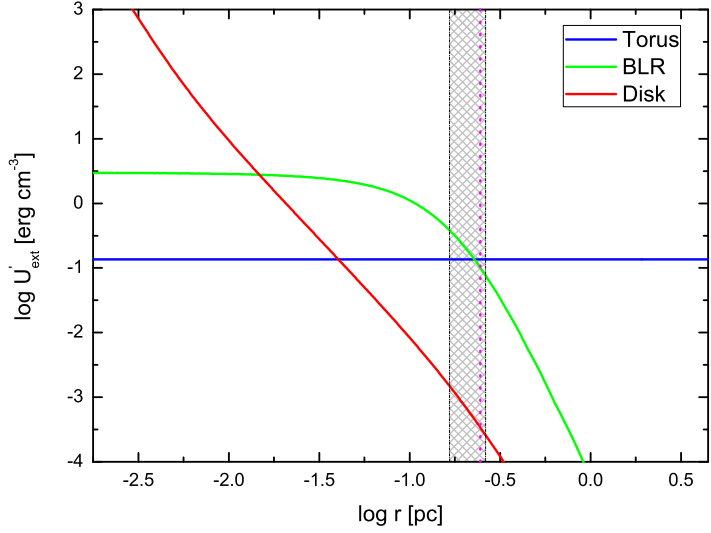}}
  \caption{Comoving energy density $U_{\rm ext}^{\prime}$ of the accretion disk, flat BLR and IR dusty torus radiation fields versus the radial distance $r$ from the BH. The shaded region is surrounded by inner radius $r_{\rm BLR,in}$ and outer radius $r_{\rm BLR,out}$ of the BLR. The vertical dashed line within the shaded region represents the location of the emission regions during both flaring states.}
\label{fig:Colored-ur}
\end{figure}

The energy densities of the external radiation fields in comoving frame of the relativistic jet as a function of radial distance $r$ from the BH are shown in
Figure \ref{fig:Colored-ur}. Close to the BH, at $r\lesssim 0.05$\, pc, the radiation field from accretion disk is the main target source for ERC process. As the distance increases, the energy density of the accretion disk rapidly decreases. At larger distance the BLR starts to dominate the cooling of electrons, and up to outer radius of the BLR, i.e., $r=r_{\rm BLR,out}\simeq 0.26$\,pc. At $r> r_{\rm BLR,out}$ the radiation field from dusty torus dominates the electron cooling. Thus the dominant radiation field responsible for the electron cooling depends on the distance of the emitting region from the BH. For 4C +21.35 the emitting regions are within the BLR clouds, the BLR photons will play a dominant role than the dusty torus radiation.

In addition to external radiation fields, the synchrotron radiation also cools electrons. The comoving energy density of the synchrotron radiation is related to the apparent isotropic synchrotron luminosity via
$u_{\rm syn}^{\prime} \simeq L_{\rm syn}^{\rm iso}/(4\pi {R_{\rm b}^{\prime}}^{2}c \delta_{\rm D}^{2})\simeq 5.33$\,erg~cm$^{-3}$, larger than $u_{\rm BLR}^{\prime}\simeq 2.63$\,erg~cm$^{-3}$, which is the energy density of the BLR at $r\simeq 0.25$\,pc where the flaring states take place. This indicates that the synchrotron radiation dominates electron cooling even though the BLR radiation is Doppler-boosted in the comoving frame.

\section{Discussion}
The discovery of the VHE emission between $\simeq 70$ and 400\,GeV with flux-doubling time about 10\,minutes from 4C +21.35 is very important to limit the location of the emitting region and the structure of the relativistic jet. The 10\,minutes variability constrains the size of the emitting region to be $R_{\rm b}^{\prime}\simeq 5\times 10^{-2}[\delta_{\rm D}/30]$\,pc. If the size is related to the cross section of relativistic jet with a half-opening angle $\sim 0.1$, the location of the emitting region well within the BLR cavity, implying that the jet of the 4C +21.35 is stratified, in which the compact emitting region is only a portion of the spine.

Observational investigations based on a substantial sample of blazars using VLBI imaging suggest that the bulk Lorentz factor and the half-opening angle of the relativistic jet satisfy $\Gamma_{\rm j}\theta_{\rm j}\sim 0.1-0.2$ \citep{2009A&A...507L..33P,2013A&A...558A.144C}. Theoretically, numerical simulations of acceleration and collimation of external-pressure-supported relativistic jets also find that $\Gamma_{\rm j}\theta_{\rm j}\lesssim 1$ \citep{2009MNRAS.394.1182K,2010NewA...15..749T}. Therefore $\theta_{\rm j}\simeq 0.2/\Gamma_{\rm j}$ is adopted in our SED modeling, while $\Gamma_{\rm j}\simeq 20$ from apparent superluminal measurement \citep{2001ApJ...556..738J}, this results in $\theta_{\rm j}\simeq 0.6$. Our modeling finally determines the location of the emitting region at $r_{\rm b}=8R_{\rm b}^{\prime}/\tan\theta_{\rm j}$. Actually, we can obtain $r_{\rm b}=4R_{\rm b}^{\prime}/\tan\theta_{\rm j}$ through the relation $\theta_{\rm j}\simeq 0.1/\Gamma_{\rm j}$. This fine-turning will increase the value of the Doppler factor and slightly decrease the normalization factor of the electron distribution, but has no significant influence on other parameters. Thus we can use the commonly relation $\theta_{\rm j}\simeq (0.1-0.2)/\Gamma_{\rm j}$ to constrain the half-opening angle of the jet of 4C +21.35, which is within the range $0.29^{\circ} \lesssim \theta_{\rm j} \lesssim 0.6^{\circ}$, implying that the emitting regions locate around the outer radius of the BLR.

\begin{figure}
  \centerline{
  \FigureFile(120mm,80mm){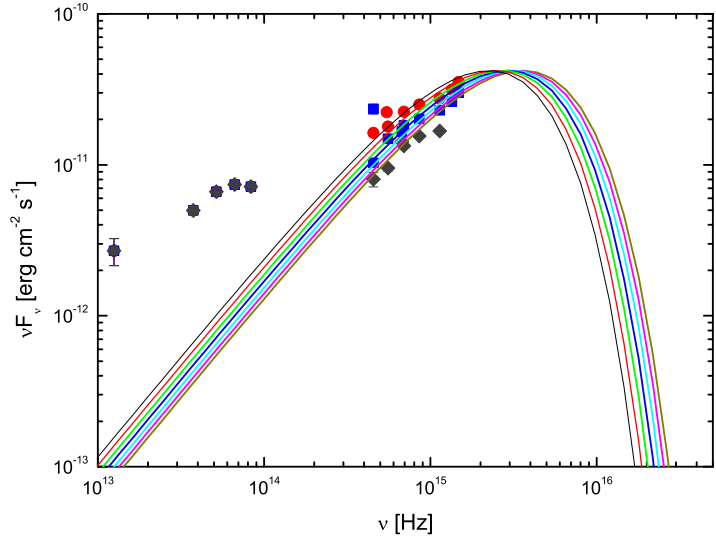}}
  \caption{Similar to Figure \ref{fig:SED-111}, but zoomed in optical and ultraviolet portion of three SEDs, which in our model originates mainly from a Shakura-Sunyaev disk emission. Model disk emission for several peak energies $E_{\rm disk,max}$ between 10\,eV and 16\,eV (with increment of unity) from left to right are shown.}
\label{fig:Colored-disk}
\end{figure}

In the optical/ultraviolet band, the clear hard spectrum and the absence of the long-term flux variations favor the dominance of a thermal component originating from the accretion disk. We thus model these data points with a multi-colored Shakura-Sunyaev disk, assuming $M_{\rm BH}=6\times 10^{8}$\,$M_{\odot}$ and $E_{\rm disk,max}=10$\,eV. The modeling SED to optical/ultraviolet data is shown in Figure \ref{fig:SED-111} with luminosity $L_{\rm d}=4.6\times 10^{46}$\,erg~s$^{-1}$. We note that $E_{\rm disk,max}=10$\,eV only gives a good fit for SED-3, we must adjust $E_{\rm disk,max}$ to fit SED-1 and SED-2. Figure \ref{fig:Colored-disk} shows an exclusive modeling to optical/ultraviolet data by varying the peak energy of the disk from 10\,eV to 16\,eV. From the statistical and systematic uncertainties, the spectrum of disk could be described by peak energy $E_{\rm disk,max}=14$\,eV.

Generally, the maximum temperature of the accretion disk, corresponding to the maximum photon energy, is assumed to be at the innermost radius. We can calculate the innermost radius with the peak energy during SED modeling as
\begin{eqnarray}
\frac{R_{\rm disk}^{\rm min}}{R_{\rm g}} &\simeq& 13.3\biggr[\frac{L_{\rm d,46}}{M_{8}^{2}~(E_{\rm disk,max}/10\,\rm eV)^{4}~\eta_{\rm d}/(1/12)}\biggr]^{1/3},
\label{eq:R_disr}
\end{eqnarray}
where $R_{\rm g}$ is gravitational radius and $\eta_{\rm d}=1/12$ is the efficiency of the gravitational potential energy of accreting matter transforming to the radiation. Using our modeling results that $L_{\rm d}=4.6\times 10^{46}$\,erg~s$^{-1}$ and $E_{\rm disk,max}=14$\,eV, the innermost radius is $0.22R_{\rm g}$, less than the normal innermost stable radius $6R_{\rm g}$ of the 4C +21.35, this results apparently is unrealistic. Even if we take $E_{\rm disk,max}=12$\,eV, which corresponds to
$1.86$\,$R_{\rm g}$, this results remain problematic! If $E_{\rm disk,max}=10$\,eV, as used in our all-multiwavelength SED modeling, the innermost radius is $\sim 6.7$\,$R_{\rm g}$, which agrees well with the innermost stable radius of a standard Shakura-Sunyaev disk. This argument provides a further support for the BH mass of 4C +21.35 with $M_{\rm BH}=6\times 10^{8}$\,$M_{\odot}$. On the other hand, this result also indicates that the ultraviolet data points shown in SED-1 and SED-2 are problematic due to some certain reasons, the use of these data points will lead to unrealistic innermost radius of the accretion disk.

\section{Conclusions}
4C +21.35 was identified as a VHE $\gamma$-ray emitter by MAGIC on 2010 June 17, where the source was in a period of high $\gamma$-ray activity detected by Fermi-LAT. The de-absorbed VHE spectrum smoothly connects to MeV-GeV spectrum from Fermi-LAT at $\sim 3$\,GeV. In the paper, we study the location of the emitting regions using three quasi-simultaneous SEDs, including a quiescent state of 2008 September 12 and two flaring states of 2010 April 29 and 2010 June 17. We preform the SED modeling based on homogeneous one-zone leptonic model, in which a flat BLR and a spherically symmetric dusty torus contribute external photon fields. We obtain the following results:

(i) The relativistic jet of the 4C +21.35 is conical with a half-opening angle within the range of $\theta_{\rm j}\simeq 0.29^{\circ}-0.6^{\circ}$.

(ii) The emitting regions locate within the BLR clouds and around the outer radius of the BLR during flaring states, while the emitting region locates beyond the dusty torus, $r>R_{\rm torus}\simeq 4.8$\,pc, for quiescent state.

(iii) The quiescent high-energy emission is dominated by SSC process. The high-energy emission during flaring periods is dominated by Compton scattering of the BLR and dusty torus radiation fields, which have comparable flux levels, the values of the Compton-dominance parameter are 29.5 and 28.5 for SED-2 and SED-3, respectively.

(iv) The SED modeling to optical/ultraviolet data supports that the BH mass of 4C +21.35 is $M_{\rm BH}=6\times 10^{8}$\,$M_{\odot}$.

\section*{Acknowledgments}
We acknowledge the anonymous referee for their critical comments and important suggestions that helped us
to improve the paper significantly. We sincerely thank Dr Justin Finke for sending us observations.
We acknowledge the financial supports from the National Natural Science Foundation of China 11133006, 11163006, 11173054, the Strategic Priority Research Program ¡°The Emergence of Cosmological Structures¡± of the Chinese Academy of Sciences (XDB09000000), and the Policy Research Program of Chinese Academy of Sciences (KJCX2-YW-T24).

\begin{table*}
\caption{List of the parameters used to construct the theoretical SEDs. Notes: Column [1]: Magnetic field strength in units of Gs; Column [2]: The bulk Lorentz factor of the emitting region; Column [3]: The spectral slope of the electron distribution low $\gamma_{\rm br}^{\prime}$; Column [4]: The spectral slope of the electron distribution above $\gamma_{\rm br}^{\prime}$. Column [5]-[6]: The minimum and the break energy of the electron energy distribution, respectively; Column [7]: the size of the emitting region in units of cm. Column: [8]-[9] The covering factor of the BLR for lines and continuum, respectively. Column: [10] The covering factor of the dusty torus. }
\label{tab:tab-1}
\begin{center}
\begin{tabular}{lcccccccccccccccc}
\hline
\hline
Notes &    B   & $\Gamma_{\rm j}$ & $s_{1}$  & $s_{2}$   &  $\gamma_{\rm min}^{\prime}$ &  $\gamma_{\rm br}^{\prime}$ & $R_{\rm b}^{\prime}$ &  $f_{\rm line}$ & $\tau_{\rm BLR}$ & $f_{\rm IR}$ \\
                 &  &     &           &     &   &($10^{3}$) &($10^{15}$) & & & & &\\
                 &[1]    &[2]    &[3]     &[4]     &[5]     &[6]     &[7]        &[8] &[9] &[10] \\
\hline
SED-1   &0.2   &18     &2.1       &4.2    &230       &3        &1     &$-$     &$-$     &$-$\\

SED-2    &0.15  &25     &2.2       &3.6    &230       &1.8        &1     &0.05    &0.05    &0.3 \\

SED-3   &0.15   &25     &2.2       &3.7    &230       &2        &1     &0.05    &0.05    &0.3\\
\hline
\hline
\end{tabular}
\end{center}
\label{tab:tab-1}
\end{table*}

\begin{table*}
\setlength{\tabcolsep}{0.02in}
\caption{The comoving energy density of the electrons ($U_{\rm e}^{\prime}$) and magnetic field ($U_{\rm B}^{\prime}$); The jet power in the form of bulk motion of electron ($P_{\rm e}$), proton ($P_{\rm p}$) and Poynting flux ($P_{\rm B}$); Absolute luminosity of synchrotron ($L_{\rm syn}$), SSC ($L_{\rm SSC}$) as well as Compton-scattered dusty torus ($L_{\rm EC}^{\rm IR}$) and BLR ($L_{\rm EC}^{\rm BLR}$) radiation fields; The total absolute radiation luminosity ($P_{\rm r}$) and the total jet power ($P_{\rm j}^{\rm tot}$). Energy density is in units of erg~cm$^{-3}$. Power and luminosity are in units of erg~s$^{-1}$. In calculation to the power related to protons, assuming one cold proton per relativistic electron.}
\label{tab:tab-2}
\begin{center}
\begin{tabular}{lcccccccccccccccc}
\hline
\hline
References & $U_{\rm e}^{\prime}$  & $U_{\rm B,-4}^{\prime}$  & $P_{\rm e,45}$  & $P_{\rm p,48}$  & $P_{\rm B,41}$ & $L_{\rm syn,42}$ &$L_{\rm SSC,43}$ & $L_{\rm EC,43}^{\rm IR}$ &$L_{\rm EC,42}^{\rm BLR}$ & $P_{\rm r,43}$  & $P_{\rm j,48}^{\rm tot}$   \\
\hline
SED-1  &12.867  &15.915   &0.786     &1.443   &0.972     &2.259  &4.478  &$-$     &$-$       &4.704     &1.444 \\

SED-2  &8.695   &8.952    &1.024     &1.881   &1.055     &1.539  &1.908  &1.173   &5.286     &3.763     &1.882 \\

SED-3  &8.951   &8.952    &1.055     &1.936   &1.055     &1.388  &1.580  &1.182   &5.433     &3.443     &1.937 \\
\hline
\hline
\end{tabular}
\end{center}
\label{tab:tab-2}
\end{table*}

\end{document}